\documentclass[prd,aps,preprint,epsfig,floats]{revtex4}
\usepackage{graphics}
\usepackage{epsfig}

\usepackage{graphicx}

 \preprint{MU-PP\#06-009}\preprint{SU 4252-824}

\begin{document}
\title{
An approach to permutation symmetry for the electroweak theory
}

 \author{Renata Jora $^{\it \bf a}$~\footnote[3]
{Email:cjora@phy.syr.edu}}

 \author{Salah Nasri $^{\it \bf b}$~\footnote[2]
{Email:snasri@physics.umd.edu}}

 \author{Joseph Schechter $^{\it \bf c}$~\footnote[4]
{Email:schechte@phy.syr.edu}}

 \affiliation{$^ {\bf \it a,c}$Department of Physics, Syracuse University,
Syracuse, NY 13244-1130}

\affiliation{$^b$ Department of Physics, University of Maryland,
College Park, MD
 20742-4111,}

\date{September 2006}

\begin{abstract}

   The form of the leptonic mixing matrix emerging from experiment has,
in the last few years, generated a lot of interest in the so-called
tribimaximal type. This form may be naturally associated with the
possibility of a discrete permutation symmetry ($S_3$) among the
three generations. However, trying to implement this attractive
symmetry has resulted in some problems and it seems to have fallen
out of favor. We suggest an approach in which the $S_3$ holds to
first approximation, somewhat in the manner of the old $SU(3)$
flavor symmetry of the three flavor quark model. It is shown that in the
case of the neutrino sector, a presently large experimentally allowed
region can be fairly well described in this first approximation.
 We briefly discuss the
nature of the perturbations which are the analogs of the Gell-Mann
Okubo perturbations but confine our attention for
the most part to the $S_3$ invariant model. We postulate that the
$S_3$ invariant mass spectrum consists of non zero masses for the
$(\tau,b,t)$ and zero masses for the other charged fermions but
approximately degenerate masses for the three neutrinos. The mixing
matrices are assumed to be trivial for the charged fermions but of
tribimaximal type for the neutrinos in the first approximation. It
is shown that this can be implemented by allowing complex entries
for the mass matrix
 and spontaneous breakdown of the $S_3$ invariance
of the Lagrangian.
\end{abstract}

\pacs{14.60.Pq, 12.15.F, 13.10.+q}

\maketitle
\section{Introduction}
    It is generally considered that a full understanding
of the masses
and associated mixings of the quarks and leptons
is one of the
chief unsolved problems in elementary particle physics.
While a great deal of experimental knowledge about quark masses and
 mixings has been available for quite a long time, it
is only in the last few years that very detailed
 information on lepton
masses and mixings has been found from a series of remarkable
experiments involving neutrino oscillations (See for some examples,
refs \cite{{superK},{kamland},{sno},{k2k},{gall},
{sage},{chooz},{minos}}). These results can
be expected to provide valuable clues toward the solution of this
mass and mixing problem. One such clue is the fact that the
leptonic mixing matrix is now known to be somewhat close to
what is called the "tribimaximal" form; actually a number of
 interesting discussions
\cite{{w},{fx},{HPS},{X},{HZ},{HS},{LV},{Z},{BHS},{MNY}}
 of this form
have been presented over a period of years. For
immediate convenience this form may be read from Eq.(\ref{R}) below
and it can be seen that one column has three equal entries while
another has two entries of equal magnitude with the
 third zero. Such a structure is natural in the context of
permutation symmetry since it is a characteristic one which
brings the basis of the defining representation
 of $S_3$ (the permutation
group on three objects) to the basis in which it is decomposed
into two and one dimensional irreducible
pieces. Of course, $S_3$ is natural in the context of the
 standard model of elementary particles, which contains three parallel
families of fermions. It has also been often discussed in the literature
\cite{{ps},{cfm},{fty},{mr}}.

     However, it seems that this symmetry can
 not be exact; otherwise there would be two exactly degenerate
neutrinos, which is not the case. It is possible that the
situation may be analogous to another three flavor theory- the
old $SU(3)$ model in which the three flavors turned out to
represent the  u, d and s quarks rather than the three families
we now know about. An initial $SU(3)$ flavor symmetry
 for the then unknown
strong interaction Hamiltonian
 was assumed to be
broken, according to the Gell Mann Okubo hypothesis \cite{gmo}, as:
\begin{equation}
H=H_0+H_3^3,
\label{symbreaker}
\end{equation}
 where $H_0$ is $SU(3)$ invariant
and $H_3^3$ breaks the symmetry , leaving the iso-spin subgroup invariant.
Actually, another symmetry breaking term, which breaks the iso-spin
subgroup but preserves a different $SU(2)$ subgroup is also required.
A possible analogy to the $S_3$ symmetry case might be to postulate that
the standard electroweak Lagrangian density has a decomposition like
\begin{equation}
{\cal L} = {\cal L}_0 + {\cal L}' + {\cal L}'' ,
\label{s3breaker}
\end{equation}
where, on the right hand side, the first term is assumed to be
$S_3$ invariant, the second term only preserves one $S_2$
subgroup of $S_3$ and the third term preserves a different
$S_2$ subgroup of $S_3$. There is a presumption that the first term
is the dominant one.

    In the present paper we will concentrate on the fully $S_3$ invariant
 term, ${\cal L}_0$ of Eq.(\ref{s3breaker}). We must specify
 what are the first approximation
fermion masses  and the first approximation mixing matrices
for which we are aiming. In the cases of the charged leptons and the
up and down quarks, the first approximation to the masses is conventionally
taken to mean assigning the heaviest fermion
 of each of the three families ($\tau,b,t$)
 a non-zero
mass and setting the masses of the others to zero. Considering that the
quark  CKM angles are all small it seems reasonable to consider
 the charged fermion mixing matrices be just the unit matrices
 in first approximation. On the other hand we would like to take the
tribimaximal form as the first $S_3$ invariant approximation to the
neutrino mixing matrix. The question of what is a good first
approximation to the spectrum of neutrino masses is perhaps not so clear.
We will argue that an approximately degnerate neutrino spectrum is the most
reasonable choice. Our job will be to demonstrate that the same $S_3$
symmetry can give rise to different patterns of masses and mixings
for the electrically charged fermions and the neutrinos.

    In this paper we will work completely in the framework of the
electroweak theory, postponing any possible inclusion in a grand unified
theory. We will assume only a minimal number of fermion fields so
that the neutrinos will be initially taken to be Majorana type. On the
other hand
we won't fully specify the Higgs fields in advance. We will  also
show that
the model can be modified for the case of Dirac neutrinos.

    In section II we will give the needed notation for the fermion fields
 and shall show how the permutation group, $S_3$ may act on them.
Section III first contains a discussion of the form of the neutrino mass
matrix under the assumption of $S_3$ invariance. It is shown how
it can be diagonalized by a matrix which is, up to a diagonal phase
matrix and a possible rotation in a two dimensional subspace, the
same as the tribimaximal form. Then, an argument is given that
the largest part of the solution space for the three neutrino masses
derived from the known mass differences corresponds to almost
complete degeneracy. Also in section III, we note that
this can easily be achieved by making use of the freedom to have
complex constants in the mass matrix. This results in a physical Majorana
phase in the mixing matrix when the neutrinos are introduced as Majorana
particles.

    Section IV contains a discussion of the mass matrix of
the charged leptons
when it is assumed that these particles acquire mass using just the
standard Higgs field. There is no problem with the $S_3$
invariance allowing for a mass spectrum in which only the $\tau$
lepton is massive, as desired. However, there is a problem with
getting a trivial charged lepton mixing matrix, as needed to realize
the tribimaximal form for the overall lepton mixing matrix. To
solve this problem one might imagine backing up and allowing
various row and column permutations for both the neutrino and charged
lepton mixing matrices (there are 36 x 36 possibilities) and possible
two dimensional rotations in the two dimensional degenerate subspaces
of each in the hope that the product comes out to be of tribimaximal
form. This hope is not realized. We give a simple proof that the
tribimaximal form can not be obtained in this manner.

    In section V, we point out a possible alternative treatment
of the charged lepton masses and mixings which seems able to give
in the first approximation, the
desired trivial mixing matrix and non-zero mass only for the tau
lepton. Instead of requiring the right handed charged leptons to
 transform under $S_3$ we consider them to be singlets under $S_3$
so that the relevant discrete group becomes just $S_{3L}$. To
construct an $S_3$ invariant Yukawa term with the Higgs fields now
requires that we introduce three different Higgs doublets, one associated
with each fermion family. Furthermore, the Higgs potential (which may
involve other possible Higgs fields too) should allow for
spontaneous breakdown of the $S_3$ symmetry in order to obtain the desired
mass spectrum and trivial mixing. As an application of the first
approximation model with three approximately
degenerate neutrinos we calculate, in this
section, the leptonic factor for neutrinoless beta decay. It depends
on the value of the degenerate neutrino masses and the
single Majorana phase in the model.

    The up and down quark masses and mixings are briefly discussed
in section VI. Since we want to have masses only for the b and t
quarks and trivial mixing in first approximation we use exactly the same
model for the Yukawa terms as in the case of the charged leptons, just
mentioned. Note that only the left handed quark fields, not the right
handed ones, are assigned to transform under $S_3$. In section VII,
we briefly discuss the form of the characteristic matrices which reflect
the invariance under any $S_2$ subgroup of $S_3$. These may be considered
as plausible perturbations to give higher approximations to this model.
We count the number of parameters corresponding to the sum of the
$S_3$ matrix and either one perturbation matrix or two perturbation
matrices invariant under different $S_2$ subgroups. Finally. section VIII
contains a brief summary and discussion.

\section{Permutation symmetry}

    For a general orientation it may be useful to start from the part of
the electroweak theory which contains just the minimal set of
fundamental fermions and their interactions with the $SU(2)_L$ x $U(1)$
gauge fields:
\begin{equation}
{\cal L}= -{\bar L}_a{\gamma}_{\mu}{\cal D}_{\mu}{L_a}
-{\bar e}_{Ra}\gamma_\mu{\cal D}_{\mu}e_{Ra}
-{\bar q}_a\gamma_{\mu}{\cal D}_{\mu}q_a
-{\bar u}_{Ra}\gamma_{\mu}{\cal D}_{\mu}u_{Ra}
-{\bar d}_{Ra}\gamma_{\mu}{\cal D}_{\mu}d_{Ra} + \cdots,
\label{firstL}
\end{equation}
where a=(1,2,3) are the generation indices, ${\cal D}_\mu$ are the
appropriate covariant derivatives for each term and
\begin{equation}
L_a=
\left[
\begin{array}{c}
{\rho}_a \\
e_{La}
\end{array}
\right], \hspace{.5in}
q_a=
\left[
\begin{array}{c}
u_{La} \\
d_{La}
\end{array}
\right].
\label{leftdoublets}
\end{equation}
The notation is conventional except that we are denoting $\rho_a$
as the two component neutrino field belonging to generation a.
We will need the charged current leptonic weak
interaction terms contained in this equation:
\begin{equation}
{\cal L}_{cc}=\frac{ig}{\sqrt{2}}W^{-}_{\mu}{\hat {\bar
e}}_L{\gamma}_{\mu}K{\hat \rho} +\hspace{.1in}h.c.+\cdots,
\label{Lcc}
\end{equation}
where $g$ is the weak coupling constant and $W_\mu^-$ is the charged
intermediate vector boson field.
The hatted fields correspond to mass eigenstates and are related to
the original ones as:
\begin{equation}
\rho=U{\hat \rho}, \hspace{.4in} e_L=W{\hat e}_L.
\label{UandW}
\end{equation}
Here, $U$ and $W$ are 3x3 unitary matrices. Finally the leptonic mixing
matrix is defined as
\begin{equation}
K=W^{\dagger}U.
\label{K}
\end{equation}

It is clear that the part of the Lagrangian given in Eq.(\ref{firstL}) (which
seems to be the most
securely established) possesses a global $U(3)^5$ symmetry- one $U(3)$
symmetry for each term. Subgroups of this very large group are
candidates for extra symmetries which might be imposed when adding the
Higgs interaction terms and potential. Discrete symmetries have the
feature that they do not necessarily imply the existence of new
Goldstone bosons when spontaneous breakdown to $U(1)_{EM}$ occurs.
The discrete symmetry which is staring us in the face is the
generational permutation symmetry $S_3$ and has been discussed by
many people. Of course, there are a number of possibilities
for implementing this symmetry. The simplest would be to have all the
5 $S_3$'s act in the same way on all the fermions. In the following we
will discuss other possibilities too. The specific symmetry
transformations take the form:
\begin{equation}
L_a \rightarrow S_{ab}L_b, \hspace{.3in}e_{Ra} \rightarrow S_{ab}e_{Rb},
\cdots
\label{symmtransf}
\end{equation}
where the permutation matrices $S$ are the 6 matrices:
\begin{eqnarray}
S^{(1)}&=&
\left[
\begin{array}{ccc}
1&0& 0 \\
0&1&0\\
0&0&1
\end{array}
\right],\hspace{.3in}
S^{(12)}=
\left[
\begin{array}{ccc}
0&1& 0 \\
1&0&0\\
0&0&1
\end{array}
\right],\hspace{.3in}
S^{(13)}=
\left[
\begin{array}{ccc}
0&0&1 \\
0&1&0\\
1&0&0
\end{array}
\right],
\nonumber \\
S^{(23)}&=&
\left[
\begin{array}{ccc}
1&0&0 \\
0&0&1\\
0&1&0
\end{array}
\right],\hspace{.3in}
S^{(123)}=
\left[
\begin{array}{ccc}
0&0&1 \\
1&0&0\\
0&1&0
\end{array}
\right],\hspace{.3in}
S^{(132)}=
\left[
\begin{array}{ccc}
0&1&0 \\
0&0&1\\
1&0&0
\end{array}
\right],
\label{s3matrices}
\end{eqnarray}
For our purpose here it will be sufficient to use this (defining)
reducible representation of $S_3$.

\section{Neutrino mass matrix}
    The simplest way to give masses to the three 2-component Majorana
neutrino fields
 is to add to the theory, in addition to the usual complex Higgs
doublet, a complex
Higgs triplet:
\begin{equation}
h=\left[
\begin{array}{cc}
h^+/\sqrt{2}&h^{++}\\
h^0&-h^+/\sqrt{2}
\end{array}
\right].
\label{triplethiggs}
\end{equation}
A vacuum expectation value for $h^0$ will then generate the neutrino mass
terms:
\begin{equation}
{\cal L}_{mass}= -\frac{1}{2}{\rho}^T{\sigma}_2M_{\nu}\rho
+\hspace{.1in}h.c.+ \hspace{.3in}\cdots
\label{Mnu}
\end{equation}
To begin with, $M_\nu$ is an arbitrary symmetric (but not necessarily
real) 3x3 matrix. Requiring $S_3$ invariance of the Lagrangian demands
invariance of this term under the transformation
$\rho_a{\rightarrow}S_{ab}\rho_b$, where S stands for any of the
permutation matrices in Eq. (\ref{s3matrices}). This
requires:
\begin{equation}
[S,M_{\nu}]=0,
\label{commutator}
\end{equation}
for all six matrices $S$. By explicitly evaluating the commutators we
obtain the solution:
\begin{equation}
M_\nu=\alpha
\left[
\begin{array}{ccc}
1&0&0\\
0&1&0\\
0&0&1
\end{array}
\right]+\beta
\left[
\begin{array}{ccc}
1&1&1\\
1&1&1\\
1&1&1
\end{array}
\right] \equiv \alpha {\bf 1}+\beta d .
\label{solution}
\end{equation}
$\alpha$ and $\beta$ are, in general complex numbers while $d$ is
usually called the ``democratic" matrix. We remark that the result
of Eq.(\ref{solution}) would be the same even for the case of Dirac
neutrinos, where the initial matrix is not required to be symmetric.

   The form of Eq. (\ref{Mnu}) indicates that the correct
``diagonalization" of $M_\nu$ should be:
\begin{equation}
U^TM_{\nu}U={\hat M}_\nu,
\label{correct}
\end{equation}
where ${\hat M}_\nu$ is a real, diagonal matrix and $U$ is a
unitary matrix. Let us perform this diagonalization while making contact
with the historical background of the subject. It is easy to verify
that $M_\nu$ may be brought to diagonal (but not necessarily real) form by
a real orthogonal matrix, $R$ as:
\begin{equation}
R^T(\alpha{\bf 1}+{\beta}d)R=\left[
\begin{array}{ccc}
\alpha&0&0 \\
0&\alpha+3\beta&0 \\
0&0&\alpha \\
\end{array}
\right].
\label{complexeigenvalues}
\end{equation}
The real matrix, $R$ is typically chosen as:
\begin{equation}
R= \left[
\begin{array}{ccc}
-2/\sqrt{6}&1/\sqrt{3}&0 \\
1/\sqrt{6}&1/\sqrt{3}&1/\sqrt{2} \\
1/\sqrt{6}&1/\sqrt{3}&-1/\sqrt{2}
\end{array}
\right]
\equiv
\left[
\begin{array}{ccc}
\vec{r}_1&\vec{r}_2&\vec{r}_3
\end{array}
\right],
\label{R}
\end{equation}
wherein we have explicitly shown how $R$ is constructed out
of the three eigenvectors (columns in $R$) of $M_\nu$.
In order to bring $M_{\nu}$ to real diagonal form we first define
the (Majorana-type) phases, $\sigma$ and $\tau$ as:
\begin{equation}
\alpha +3\beta=|\alpha+3\beta|e^{2i\sigma}, \hspace{.3in}
\alpha=|\alpha|e^{2i\tau}
\label{phases}
\end{equation}
and the diagonal matrix
\begin{equation}
P= \left[
\begin{array}{ccc}
e^{-i\tau}&0&0 \\
0&e^{-i\sigma}&0 \\
0&0&e^{-i\tau}
\end{array}
\right].
\label{Pmatrix}
\end{equation}
Now we note that the choice
\begin{equation}
U=RP
\label{defineU}
\end{equation}
yields, as desired, real positive neutrino masses:
\begin{equation}
U^T(\alpha{\bf 1}+{\beta}d)U= \left[
\begin{array}{ccc}
|\alpha|&0&0 \\
0&|\alpha+3\beta|&0 \\
0&0&|\alpha| \\
\end{array}
\right].
\label{realeigenvalues}
\end{equation}

   Most previous workers on models of this type have considered $M_{\nu}$
to be a real matrix, which results in a little simplification. Let us
discuss this case first. It is clear from a purely mathematical
standpoint, that the choice of matrix $R$ has a large amount of
arbitrariness. First, there is nothing a priori wrong with permuting
the three eigenvectors, $\vec{r}_i$ in Eq. (\ref{R}) in any way. Second,
there is nothing wrong with permuting the three rows of $R$ in any way.
In fact, there is a larger freedom of making an arbitrary rotation
in the two dimensional subspace of the degenerate eigenvectors
$\vec{r}_1$ and $\vec{r}_3$. This arbitrariness leaves $\vec{r}_2$
invariant up to its column placement. This structure is a reflection
of the possibility of decomposing the three dimensional
representation of $S_3$
into its irreducible two and  one dimensional pieces. To illustrate the
rotational freedom, we may use, instead of $R$,
\begin{equation}
R^\prime\equiv
\left[
\begin{array}{ccc}
\vec{s}_1&\vec{r}_2&\vec{s}_3
\end{array}
\right] =
\left[
\begin{array}{ccc}
\vec{r}_1&\vec{r}_2&\vec{r}_3
\end{array}
\right]
\left[
\begin{array}{ccc}
cos\xi&0&-sin\xi \\
0&1&0 \\
sin\xi&0&cos\xi
\end{array}
\right].
\label{rotation}
\end{equation}
Notice that $\vec{r}_2$ is left invariant under this rotation.
Furthermore, the matrix element, $R_{13}^\prime$ is no longer zero for
general $\xi$.

    The particular choice of $R$ written in Eq. (\ref{R}) is called
the ``tribimaximal" matrix. It seems to agree with present experimental
indications when it is identified with the lepton mixing matrix, $K$ in
Eq. (\ref{K}). (Of course, the identification of $R$ with $K$ requires the
assumption that $W$ is at least approximately the unit matrix.)
Especially,
 the $R_{13}$ element is known to be small
and could be zero. However this model can not be exactly correct because
it leads to two degenerate neutrino masses, which is ruled out since
both $m_2^2-m_1^2$ (from solar neutrino experiments) and $m_3^2-m_2^2$
(from atmospheric neutrino experiments) are non-zero and not equal to each
other.
 Thus the
$S_3$ symmetry can only be a kind of first approximation to Nature.
Nevertheless it may be a good first approximation and hence interesting to
study in detail. That is the point of view we will take here.

   In \cite{HS}, Harrison and Scott have argued that, even if one accepts
the predicted two-fold degeneracy, the $S_3$ model is still not a good
first
approximation because the two degenerate masses belong to columns 1 and 3
rather than columns 1 and 2. Masses 1 and 2 are expected to be closer
to each other because the solar neutrino mass difference is a lot smaller
than the atmospheric neutrino mass difference. Here we would like to point
out that this conclusion can be altered if we allow for complex values of
$\alpha$ and $\beta$. In that case, we can easily achieve a situation in
which all three neutrinos are degenerate. This is a valid first
approximation to the neutrino spectrum for a relatively large range of
neutrino masses. One may notice this fact by adopting a typical choice
\cite{numassdifferences} of best fit neutrino mass differences:
\begin{eqnarray}
 A \equiv m_2^2-m_1^2 &=& 6.9 \times 10^{-5} eV^2, \nonumber \\
 B \equiv |m_3^2-m_2^2| &=& 2.6 \times 10^{-3} eV^2.
\label{massdifferences}
\end{eqnarray}
The uncertainty in these numbers is roughly 25 per cent. From these
two numbers we may-with a two fold ambiguity due to the unknown sign of
$(m_3^2-m_2^2)$- determine masses $m_1$ and $m_2$ for a given choice of
$m_3$. We
denote the type I solution to be the case when $m_3$ is greater than
$m_1$ and $m_2$ and the type II solution to be the case when $m_3$
is less than $m_1$ and $m_2$.
 Note
that there is an upper limit on $m_3$ corresponding to the
cosmological
bound from structure formation \cite{cosmobound}:
\begin{equation}
m_1+m_2+m_3< 0.7~eV.
\label{cosbound}
\end{equation}
 The results are shown as Table \ref{typicalsolutions},
 parts of which have already appeared in refs. \cite{nsm}.
It should be remarked that the number of digits given for
 the neutrino
masses are only for comparing them with each other.
 They clearly do not
reflect the uncertainties of the experimental
 determinations in Eq.(\ref{massdifferences}).
\begin{table}[htbp]
\begin{center}
\begin{tabular}{cccc}
\hline \hline type & $m_1$(eV)&$m_2$(eV)&$m_3$(eV)
 \\
\hline
\hline
I & 0.2955& 0.2956& 0.3
\\
II & 0.3042& 0.3043& 0.3
\\
I & 0.2446 & 0.2447 & 0.25
\\
II & 0.2550 & 0.2551 & 0.25
\\
I & 0.1932 & 0.1934 & 0.2
\\
II & 0.2062 & 0.2064 & 0.2
\\
I & 0.1408 & 0.1411 & 0.15
\\
II & 0.1582 & 0.1584 & 0.15
\\
I & 0.0856& 0.0860& 0.1
\\
II & 0.1119& 0.1123& 0.1
\\
I & 0.0305& 0.0316& 0.06
\\
II & 0.0783& 0.0787& 0.06
\\
I & 0.0000& 0.0083& 0.0517
\\
II & 0.0643& 0.0648& 0.04
  \\
II & 0.0541& 0.0548& 0.02
 \\
II & 0.0506& 0.0512& 0.005
  \\
II & 0.0503& 0.0510& 0.001
 \\
\hline
\hline
\end{tabular}
\end{center}
\caption[]{Typical solutions for $(m_1, m_2)$
as $m_3$ is lowered from about the
highest value which is experimentally reasonable.
 In the type I
solutions $m_3$ is the largest mass
while in the type II solutions $m_3$ is the smallest mass.}
\label{typicalsolutions}
\end{table}

 In the large range 0.3 eV $>m_3>$ 0.15eV, degeneracy of the three
neutrino masses is a good first approximation to the spectrum. It is
clearly less good in the lower range  0.1 eV $>m_3>$ 0.001 eV.
Incidentally,
it is seen that only in the small range just above $m_3=$0.0517 eV
does one have anything like a usual hierarchy. An amusing
feature is that the viability of $S_3$ symmetry as a good first order
approximation selects a range of favored neutrino masses.

\begin{figure}[htbp]
\centering
\rotatebox{0}
{\includegraphics[width=10cm,height=10cm,clip=true]{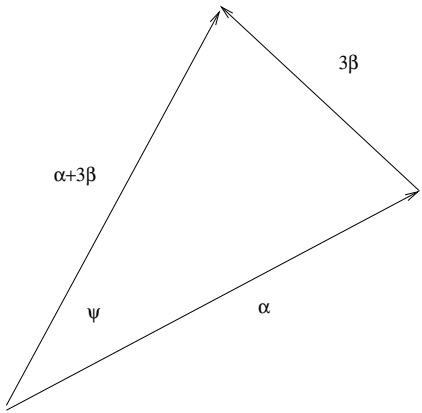}}
\caption[]{Isosceles triangle representing
 Eq.(\ref{degeneracycondition}) in the $\epsilon=0$ limit.}
\label{triangle}
\end{figure}

    Actually, at first glance it may not be apparent how
 Eq. (\ref{realeigenvalues}) allows for a degenerate mass spectrum. To
achieve an approximately degenerate mass spectrum
we impose:
\begin{equation}
|\alpha|=|\alpha+3\beta| +\epsilon,
\label{degeneracycondition}
\end{equation}
where $\epsilon$ is a real number which may be as small as
one likes.
A picture of this equation with $\epsilon=0$, where $\alpha$ and
$\alpha+3\beta$ are displayed as 2 dimensional vectors
making up the equal sides of an isosceles triangle is given
in Fig.\ref{triangle}. The angle, $\psi=2(\sigma-\tau)$ between them must
obey (to first order in $\epsilon$):
\begin{equation}
sin(\frac{\psi}{2})=\frac{3|\beta|}{2|\alpha|}
(1+\frac{\epsilon}{2|\alpha|}).
\label{psi}
\end{equation}

It is interesting to contrast this solution with
one in which the coefficient $\beta$ in Eq.(\ref{solution})
vanishes \cite{{fty},{mr}}.
Then the three neutrino masses are obviously degenerate but $R$ in
Eq.(\ref{complexeigenvalues}) can be any orthogonal matrix. However,
in the present case, for any finite  $\epsilon$ no matter how small,
 $R$ can have at most a two dimensional degenerate
subspace.We will see that there is a physical effect which
persists for small $\epsilon$.

 Since only experiment can tell us eventually whether neutrinos
are of Majorana or Dirac type, it seems worthwhile to show that the same
results may be obtained in case the neutrinos are of Dirac type. In that
case we have the two fields $\nu_L\equiv\rho$ and $\nu_R$ so the mass term
becomes $-{\bar \nu}_RM_{\nu}\nu_L +h.c.$. The diagonalization equation
which replaces Eq.(\ref{correct}) is:
\begin{equation}
U_R^{\dagger}N_{\nu}U_L={\hat M}_{\nu}.
\label{alternate}
\end{equation}
If, for example, both $\nu_L$ and $\nu_R$ transform as triplets
under the same $S_3$, we will again have the same form of $M_{\nu}$
as given in Eq. (\ref{solution}). The bi-diagonalizing matrices then
become
\begin{equation}
U_L=RP, \hspace{.3in} U_R=RP^*,
\label{Diracdiag}
\end{equation}
where $R$ is given in Eq.(\ref{R})
and $P$ is given in Eq.(\ref{Pmatrix}).
This replaces Eq.(\ref{defineU}) so
the neutrino contribution to the overall lepton
 mixing matrix is just the
same. In this case, however, the extra phases, $P$ in $U_L$ may be
transformed away. There is no Majorana phase and, of
 course, no neutrinoless double
beta decay. Nevertheless, the complex phase plays a role in the
mass eigenvalues, allowing all three neutrinos to be degenerate
 with non zero $\beta$.

\section{Charged lepton mass matrix}
    The charged leptons may simply obtain their masses using the
conventional complex doublet Higgs field,
\begin{equation}
\Phi=\left[
\begin{array}{c}
\phi^+ \\
\phi^0
\end{array}
\right],
\label{usualhiggs}
\end{equation}
in the Lagrangian term,
\begin{equation}
{\cal L} =-\frac{\sqrt{2}}{v}{\bar e}_{Ra}(M_e)_{ab}\Phi^{\dagger}L_b +
h.c.
+\cdots,
\label{electronmassterm1}
\end{equation}
where $v=\sqrt{2}<\phi^0>$ and $\phi^0=(v+{\tilde {\phi}}^0)/\sqrt{2}$.
Here,
 $M_e$ is the pre-diagonal charged lepton mass matrix. This matrix may, in
general, be brought to a real diagonal form ${\hat M}_e$ by the
bi-unitary transformation,
\begin{equation}
M_e=U_e{\hat M}_eW^\dagger,
\label{biunitary}
\end{equation}
where,
\begin{equation}
M_e^{\dagger}M_e=W{\hat M}_e^2W^{\dagger}, \hspace{.3in}
M_eM_e^{\dagger}=U_e{\hat M}_e^2U_e^{\dagger}.
\label{bidiagonalization}
\end{equation}
This leads to the mass diagonal, hatted, fields:
\begin{equation}
e_L=W{\hat e}_L, \hspace{.3in} e_R=U_e{\hat e}_R.
\label{elefthat}
\end{equation}

    So far we haven't discussed the imposition of $S_3$ symmetry on
$M_e$. Because the fields $\rho_a$ and $e_{La}$ both belong to the same
$SU(2)_L$ doublet we should, to maintain the $SU(2)_L$ symmetry, require
$e_{La}$ to transform in the same manner as $\rho_a$. However $e_{Ra}$
is not required to do so. Nevertheless that is the obvious first
guess so let us initially require invariance of the Lagrangian under
both transformations in Eq. (\ref{symmtransf}). Then we must also impose
\begin{equation}
[S,M_e]=0,
\label{Meform}
\end{equation}
for all six permutation matrices.
As before, this has the solution,
\begin{equation}
M_e=\gamma{\bf 1}+{\delta}d,
\label{solveMe}
\end{equation}
where $d$ is the previously defined democratic matrix while $\gamma$
and $\delta$ are two complex numbers.
${\hat M}_e$ will then have have the doubly degenerate
real positive eigenvalue $|\gamma|$ as well
as the non degenerate eigenvalue $|\gamma+3\delta|$. We define
phases $\sigma'$ and $\tau'$ by,
\begin{equation}
\gamma + 3\delta = |\gamma + 3\delta|e^{2i\sigma'}
\hspace{.3in}
\gamma=|\gamma|e^{2i\tau'}.
\label{morephases}
\end{equation}
If these phases are used to define a diagonal matrix, $P'$ of
the form,
\begin{equation}
P'=\left[
\begin{array}{ccc}
e^{-i\tau'}&0& \\
0&e^{-i\sigma'}&0 \\
0&0&e^{-i\tau'}
\end{array}
\right]
\label{Pprime}
\end{equation}
we can satisfy Eq.(\ref{biunitary}) with the identifications,
\begin{equation}
W=R'P' \hspace{.3in}
U_e=R'P'^*,
\label{WandUe}
\end{equation}
where $R'$ is a real orthogonal matrix of the
generalized (as discussed around Eq. (\ref{rotation}))
tribimaximal type. We remark that it is not necessary for the
ordering of the eigenvalues in ${\hat M}_e$ (and correspondingly
in Eq.(\ref{Pprime}) to be the same as in ${\hat M}_\nu$.
As discussed above, we can also permute the rows of $R'$
and make an arbitrary rotation in the degenerate subspace.
A reasonable first approximation to the charged lepton mass
spectrum would seem to consist of a single massive
 $\tau^-$ and two massless
(or small mass) others ($e^-$ and $\mu^-$). This can easily be
achieved by setting $\gamma$ to zero and identifying the
third eigenvalue to be the non-degenerate one.

   However, it is not so easy to get the correct lepton mixing matrix
which, using Eq.(\ref{K}), takes the form,
\begin{equation}
K=P'^*R'^T{RP}.
\label{firsttryK}
\end{equation}
Note that the phase matrix $P'^*$ can be eliminated by a rephasing
of the field ${\hat {\bar e}}_L$, which sits to the left of it in
the physical interaction term, Eq.(\ref{Lcc}). The question is whether
the product $R'^T{R}$ can have anything like the tribimaximal form.
Ideally, one might like to keep $R$ as the tribimaximal matrix and
arrange for $R'$ to approximate the unit matrix. This could be
achieved if the complex number $\delta$ in Eq. (\ref{solveMe}) could be
made zero.
However we have seen that it is necessary for $\gamma$ to
 be zero in order
to give a realistic first approximation to the charged lepton
mass spectrum. Clearly, having $\gamma=\delta=0$ is not
viable.
 Then, $R'$ can not be the unit matrix but
 must be of generalized tribimaximal
type. It is not expected that $R'$ would be the same as $R$ since we have
seen that a different ordering of mass eigenvalues is appropriate. In
general
there are 36 x 36 discrete possibilities for the matrix $R'^TR$
corresponding to 6 possible row exchanges and six possible column
exchanges for each and also
the possibilities of two independent rotations in the two degenerate
subspaces. At first, it seems to be a daunting task to study them.
Nevertheless we will demonstrate a simple way to see that there are
no acceptable solutions that would give a resulting $K$ anything like
the tribimaximal form desired.

    For this task it is convenient to represent each of the relevant
 matrices as a single row, having elements which are column vectors as in
Eqs. (\ref{R}) and (\ref{rotation}). To anchor the notation we introduce
unit vectors:
\begin{equation}
\vec{n}_1=\left[
\begin{array}{c}
1\\0\\0
\end{array}
\right],
\hspace{.15in}
\vec{n}_2=\left[
\begin{array}{c}
0\\1\\0
\end{array}
\right],
\hspace{.15in}
\vec{n}_3=\left[
\begin{array}{c}
0\\0\\1
\end{array}
\right].
\label{unitvectors}
\end{equation}
Now the matrices $R=[\vec{r}_1,\vec{r}_2,\vec{r}_3]$ and
 $R'=[\vec{t}_1,\vec{t}_2,\vec{t}_3]$ may be written as row
vectors with indexed column vector elements according to:
\begin{equation}
\vec{r}_a=\vec{n}_bR_{ba}, \hspace{.3in}\vec{t}_a=\vec{n}_bR'_{ba}.
\label{compactnotation}
\end{equation}
From the first of these equations we may write $\vec{n}_c=\vec{r}_a
R_{ac}^{-1}$. Substituting this into the second of Eqs.
(\ref{compactnotation}) we find $\vec{t}_f=\vec{r}_aR_{ac}^{-1}R'_{cf}$,
which may be more compactly written as:
\begin{equation}
\vec{t}_f=\vec{r}_a{\tilde K}_{af}.
\label{Ktilde}
\end{equation}
Here ${\tilde K}=KP^*$ is the non-phase part of the lepton mixing matrix.
In other words, the lepton mixing matrix is simply displayed as
the transformation between the columns of $R$ and $R'$. The key point
for our present purpose is that one of the $\vec{t}_f$ and one of
the $\vec{r}_a$ must be:
\begin{equation}
\frac{1}{\sqrt{3}}\left[
\begin{array}{c}
1 \\ 1 \\ 1
\end{array}
\right].
\label{invariantvector}
\end{equation}
This is seen to be the case for any generalized
tribimaximal matrix which diagonalizes permutation
 invariant mass matrices in Eq. (\ref{rotation}).
It is also clear that an arbitrary permutation of rows
in the diagonalizing matrix does not change this vector.
Without any loss of generality let the invariant vector in
Eq.(\ref{invariantvector}) occur as $\vec{t}_1$ and $\vec{r}_2$.
Then Eq.(\ref{Ktilde}) becomes,
\begin{equation}
\vec{t}_1=\vec{r}_1{\tilde K}_{11}+\vec{t}_1{\tilde
K}_{21}+\vec{r}_3{\tilde K}_{31}.
\label{condition}
\end{equation}
Since the three vectors on the right hand side are linearly
independent, we conclude that ${\tilde K}_{21}=1$ and in addition
${\tilde K}_{11}={\tilde K}_{31}=0$. Similarly, the inverse relation
$\vec{r}_b=\vec{t_c}{\tilde K}_{cb}^{-1}$ yields  again
${\tilde K}_{21}=1$ and also ${\tilde K}_{22}={\tilde K}_{23}=0$.
Thus the matrix ${\tilde K}$ must have the form,
\begin{equation}
{\tilde K}= \left[
\begin{array}{ccc}
0 & x & x \\
1 & 0 & 0 \\
0 & x & x
\end{array}
\right],
\label{Ksofar}
\end{equation}
where the x's stand for elements yet to be determined.
Since ${\tilde K}$ is a real orthogonal matrix we can
finally write:
\begin{equation}
{\tilde K}= \left[
\begin{array}{ccc}
0 & {\pm c} & {\pm s} \\
1 & 0 & 0 \\
0 & -s & c
\end{array}
\right],
\label{finally}
\end{equation}
where $s$ and $c$ stand for the sine and cosine of some angle.
The key point is that there are four zero elements of the matrix
and one independent parameter. Therefore it represents, up to some
possible
reflections and relabeling of axes, a rotation in a two dimensional
subspace. This
structure clearly does not depend on the choices of $\vec{r}_a$
and $\vec{t}_b$ we chose to identify with the invariant vector of
Eq.(\ref{invariantvector}).
The tribimaximal matrix of Eq.(\ref{R}) and the generalized tribimaximal
mixing matrix of Eq.(\ref{rotation}) can not be obtained from
this too simple form.

   To sum up the work so far, we have seen that $S_3$ symmetry based on
the transformations properties for $L_a$ and $e_{Ra}$
 as given in Eq.(\ref{symmtransf})
has some very encouraging features for a first approximation to lepton
properties. First, the neutrino mass spectrum is consistent with
three approximately  degenerate masses. In addition,
 the predicted neutrino mixing matrix
is consistent with the tribimaximal matrix, as experiment suggests.
Also, the charged lepton mass pattern can be made consistent with a heavy
$\tau^-$ and zero mass $\mu^-$ and $e^-$. Things would be fine if the
charged lepton mixing matrix could be made consistent with the unit matrix
so that, as in Eq.(\ref{K}), the lepton mixing matrix, $K$ could be
identified with the tribimaximal form (up to the possibility of Majorana
 phases). Unfortunately, we have just shown that this is not possible.
We will now investigate a modification of the field transformation
properties which looks able to solve this problem.

For clarification we remark that the problem above would not have
arisen if we had not assumed the restriction on the $S_3$ invariant
form of the charged lepton spectrum that only the $\tau$ particle
has mass. If we could tolerate a first approximation in which
$(e,\mu,\tau)$ are all degenerate, then the choice of
 $\delta=0$ in Eq.(\ref{solveMe}) would work. However that does not seem
physically reasonable.

\section{Modified approach for charged leptons}
     As mentioned around Eq.(\ref{symmtransf}), there is some freedom to
modify
 the type of the $S_3$ transformations we assume. In the last
section we assumed that $L_a$ and $e_{Ra}$ transform in the same way
so we were identifying the $S_{3L}$ and $S_{3R}$ transformations.
Now let us assume that $e_{Ra}$ does not transform under $S_{3L}$; i.e.,
it transforms as a singlet under $S_{3L}$. Then, in order to construct an
invariant charged lepton mass term we need to assume that there are
three Higgs doublets which belong to the (reducible) defining
representations of $S_{3L}$:
\begin{equation}
\Phi_a=\left[
\begin{array}{c}
\phi_a^+ \\
 \phi_a^0
\end{array}
\right],
\label{phia}
\end{equation}
where $\phi_a^0=(v_a+{\tilde \phi}_a^0)/\sqrt{2}$. The charged lepton
masses may then arise from a term,
\begin{equation}
{\cal L}=-\sqrt{2}\sum_{c,b}{\bar
e}_{Rc}\sum_a\Phi_a^{\dagger}G_{ab}^{(c)}L_b + \hspace{.15in} h.c.,
\label{3phiterm}
\end{equation}
where $G_{ab}^{(c)}$ is a matrix of coupling constants whose form
will be restricted by the $S_{3L}$ symmetry (The $S_{3R}$ symmetry is not
being implemented now). The charged lepton mass matrix is:
\begin{equation}
(M_e)_{cb}=\sum_{a}v_aG_{ab}^{(c)},
\label{3phiMe}
\end{equation}
so that the charged lepton mass term in the Lagrangian looks
 conventional:
\begin{equation}
{\cal L}= -\sum_{c,b}{\bar e}_{Rc}(M_e)_{cb}e_{Lb}+\hspace{.1in} h.c.
\label{usualMe}
\end{equation}
As before, the $S_{3L}$ invariance demands that the three matrices,
$G^{(c)}$ have the forms:
\begin{equation}
G^{(c)}= \gamma^{(c)}{\bf 1}+\delta^{(c)}{d},
\label{Gc}
\end{equation}
wherein the six quantities $\gamma^{(c)}$ and $\delta^{(c)}$ are
all undetermined complex numbers. Then the predicted form for
$M_e$ is:
\begin{equation}
M_e=\left[
\begin{array}{ccc}
v_1\gamma^{(1)}+\lambda\delta^{(1)}&v_2\gamma^{(1)}+\lambda\delta^{(1)}&
v_3\gamma^{(1)}+\lambda\delta^{(1)} \\
v_1\gamma^{(2)}+\lambda\delta^{(2)}&v_2\gamma^{(2)}+\lambda\delta^{(2)}&
v_3\gamma^{(2)}+\lambda\delta^{(2)} \\
v_1\gamma^{(3)}+\lambda\delta^{(3)}&v_2\gamma^{(3)}+\lambda\delta^{(3)}&
v_3\gamma^{(3)}+\lambda\delta^{(3)}
\end{array}
\right],
\label{bigMe}
\end{equation}
where we have denoted the sum of the three vacuum values as
$\lambda=v_1+v_2+v_3$.

    Our goal is to see if Eq.(\ref{bigMe}) can yield a mass spectrum with
a single massive charged lepton and the unit matrix for $W$, the charged
lepton mixing matrix defined in Eq.(\ref{UandW}). This can be simply
achieved if
first, the three arbitrary constants $\delta^{(a)}$ are all set to zero
together with the two arbitrary constants $\gamma^{(1)}$ and
$\gamma^{(2)}$. Second, since just the bottom row remains, we must have
a spontaneous breakdown structure in the Higgs potential which results
in
the vanishing of the expectation values $v_1$ and $v_2$ but
 not $v_3$. The $\tau^-$
mass then becomes, in the first approximation limit, $v_3\gamma^{(3)}$.

       The Higgs potential appropriate for this model may be rather
complicated since many additional Higgs fields could very well be present.
One characteristic term illustrating how a basic permutation invariant
quadratic form appears is:
\begin{equation}
V=k_1\left[\Phi_a^{\dagger}(\epsilon\delta_{ab}
+\zeta{d_{ab}})\Phi_b-k_2\right]
\left[\Phi_e^{\dagger}(\epsilon\delta_{ef}
+\zeta{d_{ef}})\Phi_f-k_2\right]^*+\hspace{.15in} \cdots,
\label{higgsv}
\end{equation}
where $k_1$,$k_2$,$\epsilon$ and $\zeta$ are constants while
$d_{ab}$ stands for elements of the democratic matrix.

   So it seems that by using, instead of a single Higgs doublet,
three Higgs doublets linked to the three leptonic families by
$S_{3L}$ invariance, we can obtain a suitable first approximation to the
lepton mass spectrum and to the leptonic mixing matrix ${K}$.
Specifically, the charged leptons have a trivial mixing matrix and only
the heaviest one is massive. On the other hand, in this same
approximation,
the three neutrino masses are approximately degenerate
 and have a mixing matrix
which can be taken to be of the tribimaximal form, Eq.(\ref{defineU})
or its generalization to include a rotation in the two
dimensional degenerate subspace.

 Associated
with this model in the case where
the neutrinos are formulated as Majorana particles, is the existence of
the Majorana
phase $\psi$
defined in Eq.(\ref{psi}). To illustrate the importance
of this CP violating Majorana phase we calculate the characteristic
leptonic factor $|m_{ee}|$ for neutrinoless double beta decay in the
present model with three approximately degenerate neutrinos
 having mass, $m=|\alpha|$.
The general formula is:
\begin{equation}
|m_{ee}|=|m_1(K_{11})^2+m_2(K_{12})^2+m_3(K_{13})^2|,
\label{generalmee}
\end{equation}
where the lepton mixing matrix elements are obtained by identifying
$K$ with $U$ given in Eqs.(\ref{defineU}), (\ref{Pmatrix}) and
(\ref{R}). This yields with $\epsilon$ negligible,
\begin{equation}
|m_{ee}|=\frac{m}{3}\sqrt{5+4cos{\psi}}.
\label{ourmee}
\end{equation}
It is easy to check that this result would not change if we computed
the matrix elements $K_{ab}$ using the rotated form of $R$ given
in Eq.(\ref{rotation}) rather than the one in Eq.(\ref{R}).
Clearly $|m_{ee}|$ obeys the inequality,
\begin{equation}
m\geq{|m_{ee}|}\geq{m/3}.
\label{inequality}
\end{equation}
The present experimental bound on this
quantity is \cite{kka},
\begin{equation}
|m_{ee}|<(0.35-1.30)eV,
\label{expbound}
\end{equation}
which is not much greater
than the range of m for which the
approximate degeneracy of all three
neutrinos holds. Thus this kind of model may
be tested in the next few years.

\section{Quarks}
      The up and down quark mass matrices clearly have a reasonable first
approximation in which the heaviest of each is massive while the other
two have no mass. In the same approximation it is reasonable to have a
unit mixing matrix. Then the situation for each is the same as for the
charged leptons, as just discussed. It thus seems natural to use the same
setup
as for the charged leptons, with three Higgs fields transforming
under $S_{3L}$. The quark mass terms in the Lagrangian
 would take the
form,
\begin{equation}
{\cal L}=-\sqrt{2}\sum_{c,b}{\bar
d}_{Rc}\sum_a\Phi_a^{\dagger}B_{ab}^{(c)}q_b +
i\sqrt{2}\sum_{c,b}{\bar
u}_{Rc}\sum_a\Phi_a^TA_{ab}^{(c)}\tau_2q_b + \hspace{.15in} h.c.,
\label{quarkhiggs}
\end{equation}
which yields the up and down quark mass matrices as:
\begin{equation}
(M_d)_{cb}=\sum_{a}v_aB_{ab}^{(c)}, \hspace{.2in}
(M_u)_{cb}=\sum_{a}v_aA_{ab}^{(c)}.
\label{quarkmasses}
\end{equation}
The predictions of $S_3$ invariance are
\begin{equation}
A^{(c)}= \eta^{(c)}{\bf 1}+\theta^{(c)}{d},\hspace{.2in}
B^{(c)}= \iota^{(c)}{\bf 1}+\kappa^{(c)}{d}.
\label{AandB}
\end{equation}
As in the charged lepton case the first approximation
 to the quark mass spectrum yields,
\begin{equation}
m_t=v_3\eta^{(3)}, \hspace{.2in} m_b=v_3\iota^{(3)},
\label{tbmasses}
\end{equation}
while the other quark masses remain zero.
For completeness we mention that the up and down
mass matrices in the general case are brought to
diagonal forms by the bi-unitary transformations:
\begin{equation}
M_u=U_u{\hat M}_uW^{\dagger}_u, \hspace{.2in}
M_d=U_d{\hat M}_dW^{\dagger}_d.
\label{quarkbiunitary}
\end{equation}
This leads to the diagonal (hatted) states,
\begin{equation}
u_L=W_u{\hat u}_L, \hspace{.15in}
u_R=U_u{\hat u}_R, \hspace{.15in}
d_L=W_d{\hat d}_L, \hspace{.15in}
d_R=U_d{\hat d}_R,
\label{hattedquarks}
\end{equation}
and the hadronic part of the charged current
weak interaction,
\begin{equation}
{\cal L}_{cc}=\frac{ig}{\sqrt{2}}W^{+}_{\mu}{\hat {\bar u}}_L
{\gamma}_{\mu}C{\hat d}_L +\hspace{.1in}h.c.+\cdots.
\label{C}
\end{equation}
Here $C=W_u^{\dagger}W_d$ is the quark mixing (or CKM) matrix.
In the first approximation under consideration $C={\bf 1}$.

\section{Higher order approximations}
    Next let us briefly consider the perturbations which
 might be employed to improve the first approximation fermion
spectra and mixing matrices. For this purpose, we first summarize
the general form of the $S_3$ invariant matrices which appear in
Eqs. (\ref{solution}), (\ref{solveMe}), (\ref{Gc}) and
 (\ref{AandB}) as:
\begin{equation}
\left[
\begin{array}{ccc}
a&b&b \\
b&a&b \\
b&b&a
\end{array}
\right].
\label{s3type}
\end{equation}
This matrix contains two complex parameters. As we have
seen the non-reality plays an important role in obtaining
the proposed first approximation to the neutrino mass
spectrum. Of course, the parameters will be different
for fermions of each electric charge. The assumption of a grand
unified theory would give additional relations among them.
Note also that the $S_3$ invariant matrix and possible
perturbations of it directly correspond to the the mass matrix for
the assumed Majorana neutrinos but not for the charged leptons and quarks.
In the latter cases, the effects of the three Higgs mesons being assumed
must be folded in as in Eqs.(\ref{3phiMe}) and
 (\ref{quarkmasses}). An additional difference is that the matrix
for the Majorana neutrino masses must be a symmetric one.

    If we assume that the perturbations are invariant under
only an $S_2$ subgroup of $S_3$, there are three choices.
These correspond to two element subgroups containing, besides
the identity element, $S^{(12)}$, $S^{(13)}$ or $S^{(23)}$.
Requiring, a general matrix, $M$ to commute with these
separately  gives,
respectively the $S_2$ invariant forms:
\begin{equation}
\left[
\begin{array}{ccc}
a'&b'&c' \\
b'&a'&c' \\
d'&d'&e'
\end{array}
\right], \hspace{.2in}
\left[
\begin{array}{ccc}
a''&c''&b'' \\
d''&e''&d'' \\
b''&c''&a''
\end{array}
\right],   \hspace{.2in}
\left[
\begin{array}{ccc}
e'''&d'''&d''' \\
c'''&a'''&b''' \\
c'''&b'''&a'''
\end{array}
\right].
\label{s2matrices}
\end{equation}
Each of these possible perturbation matrices contains
five complex parameters. In case of application to the Majorana
neutrino masses, there will be only four complex parameters
since we have the corresponding symmetric structures:
\begin{equation}
\left[
\begin{array}{ccc}
a'&b'&c' \\
b'&a'&c' \\
c'&c'&e'
\end{array}
\right], \hspace{.2in}
\left[
\begin{array}{ccc}
a''&c''&b'' \\
c''&e''&c'' \\
b''&c''&a''
\end{array}
\right],   \hspace{.2in}
\left[
\begin{array}{ccc}
e'''&c'''&c''' \\
c'''&a'''&b''' \\
c'''&b'''&a'''
\end{array}
\right].
\label{s2majoranamatrices}
\end{equation}
    The simplest perturbation scheme would consist of the original form,
Eq.(\ref{s3type}) plus one of the three matrices in Eq.(\ref{s2matrices})
[or Eq.(\ref{s2majoranamatrices}) for the Majorana neutrino case].
Each of the three possible sums can be seen to have five [four] complex
parameters, as one can absorb the two $S_3$ invariant matrix parameters
in the parameters of the perturbing matrix. Presumably, one of
these three possibilities would be the best.

    The perturbation scheme which might be the closest analog of the
generalized Gell-Mann Okubo type mentioned in the Introduction would
contain
the $S_3$ invariant part together with two of the matrices in
 Eq.(\ref{s2matrices})[ Eq.(\ref{s2majoranamatrices})] .
 It can be seen that one would get the same
number, eight [six], of complex parameters regardless of which set of two
is selected. In fact, regardless of which set is chosen, the resulting
sum of three matrices would be the same. In the case of the Majorana
neutrino matrix, the six complex parameters already comprise the most
general matrix. For the charged fermion matrices,
 nine parameters are needed for the most general possibility; however,
all of the perturbation matrices satisfy the relation:
\begin{equation}
M_{12}+M_{23}+M_{31}=M_{32}+M_{21}+M_{13},
\label{constraint}
\end{equation}
thereby eliminating one complex parameter. Clearly, there is no point
in considering a perturbation made as the sum of all three matrices
in Eq.(\ref{s2matrices}).

   The most practical second approximation would seem to
involve choosing just one of the matrices of Eq.(\ref{s2matrices})
[Eq.(\ref{s2majoranamatrices})].  A promising choice of
 perturbation,
 where the $S_2$ subgroup includes the element $S^{(23)}$ has been
investigated by a number of authors \cite{mutau, moh, Gutmutau}.

\section{Summary and discussion}

    For a long time the permutation invariance, $S_3$  of the
three known generations has been considered a very natural
assumption for understanding the non-trivial spectrum of
fundamental fermion masses and their associated mixings.
Recently, experimental data on neutrino oscillations have
pointed to a leptonic mixing matrix which is of the
 so called tribimaximal form; this is in essence
 the transformation from
the natural three dimensional basis of $S_3$ to the
irreducible basis and therefore can be regarded as
another motivation for this group.
    We have attempted to spell out in detail a possible way
in which this attractive $S_3$ permutation invariance
 can be consistently implemented
in the standard electroweak theory.

    We started by reviewing the fact that exact $S_3$
invariance is not correct and suggesting that the situation
might be analogous to an older one where a symmetry group was very
plausible but the exact dynamics were unknown--the SU(3) symmetry
of the three quark model. A similar proposal is that the $S_3$
invariance holds for a dominant piece but that there are other (presumably
relatively small) pieces which preserve different $S_2$ subgroups
of $S_3$. In this paper we concentrated on looking at the
first approximation of exact $S_3$ invariance. We showed,
 by comparison with results of analysis of neutrino oscillation
experiments (See Table I) that there is a large range of
 allowed neutrino mass sets where near degeneracy of all
three neutrino masses is a very good approximation. So
we regarded a near degenerate neutrino mass spectrum together
with a tribimaximal type neutrino mixing matrix as the
$S_3$ invariant form to be obtained. At the same time we
considered that the $S_3$ invariant structures for the charged
leptons and quarks should contains masses for the ($\tau,t,b$)
states and zero masses for the others. The $S_3$ invariant
mixing matrices for the charged fermions
are all considered to be just the
unit matrix (no mixing to leading order).

    It is not completely trivial to obtain a model
 which achieves this desired spectrum. To get it, we
 included the features
 of i) Majorana type CP violation, ii) $S_3$ invariance
for only the left fermions, iii) Inclusion of three Higgs
doublets linked to the three generations and transforming
under $S_3$ and iv) spontaneous breakdown of $S_3$ due
to only one of those three doublets developing a vacuum
value. The motivation is given in section IV in which it
 is shown that, very generally, one can not get a tribimaximal
form for the overall lepton mixing matrix
together with the given mass spectrum of the charged leptons
 in the simplest way
one would imagine.

    A physical prediction of the model
 when the neutrinos are assumed to be of Majorana type
is obtained for the neutrinoless double beta decay
factor $m_{ee}$ already in the $S_3$ invariant limit
(See  Eq. (\ref{ourmee})). This gives a bound which
should be testable in the relatively near future if
the three neutrinos belong to the energy range where they are
approximately degenerate.

    One natural extension of this approach would be to look
at its embedding in a grand unified theory. The next step would
be to investigate in detail the case when a perturbation
 breaking the symmetry down to an $S_2$ subgroup is included.
 As discussed in Section VII, this would introduce two additional
complex parameters for the neutrino mass matrix.

\section*{Acknowledgments}
\vskip -.5cm

    We would like to thank Lincoln Wolfenstein for a helpful
comment on the manuscript and together with Samina Masood and Susumu Okubo
for helpful discussions. The work of
 R.Jora and J.Schechter has been supported in
part by the US DOE under Contract No. DE-FG-02-85ER 40231. S.N is
supported by National Science Foundation (NSF) Grant No.
PHY-0354401.


\begin{thebibliography}{9}

\bibitem{superK}Super Kamiokande collaboration, S. Fukuda et al,
Phys. Lett. B {\bf 539}, 179 (2002), hep-ex/0205075.

\bibitem{kamland}KamLAND collaboration, K. Eguchi et al, Phys. Rev.
Lett. {\bf 90}, 021802 (2003).

\bibitem{sno}SNO collaboration, Q. R. Ahmad et al,nucl-ex/
0309004.

\bibitem{k2k}K2K collaboration, M. H. Ahn et al, Phys. Rev. Lett. {\bf
90}, 041801 (2003).

\bibitem{gall}GALLEX Collaboration, W. Hampel et al, Phys. Lett. B
{\bf 447}, 127 (1999).

\bibitem{sage}SAGE Collaboration, J. N. Abdurashitov et al,
Phys. Rev. C {\bf 60}, 055801 (1999).

\bibitem{chooz}CHOOZ Collaboration, M. Apollonio et al,
Eur. Phys. J. C {\bf 27}, 331 (2003), hep-ex/0301017.

\bibitem{minos}MINOS Collaboration, Phys. Rev. D {\bf 73}, 072002
 (2005), hep-ex/0512036.

\bibitem{w}L. Wolfenstein, Phys. Rev. D {\bf 18}, 958 (1978).

\bibitem{fx}H. Fritzsch and Z.-Z.Xing, Phys. Lett. B {\bf 440},
313 (1988), hep-ph/9808272.

\bibitem{HPS}P. F. Harrison, D. H. Perkins and W. G. Scott, Phys.
Lett. {\bf B530},79 (2002), hep-ph/0202074.

\bibitem{X}Z.-Z.Xing, Phys. Lett. B {\bf 533}, 85 (2002),
 hep-ph/020409.

\bibitem{HZ}X.G.He and A. Zee, Phys. Lett. B {\bf 560}, 87 (2003),
hep-ph/0204049.

\bibitem{HS} P. F. Harrison and W. G. Scott, hep-ph/0302025.

\bibitem{LV}C.I.Low and R.R.Volkas, Phys. Rev. D {\bf 68},
033007 (2003), hep-ph/0305243.

\bibitem{Z}A.Zee, Phys. Rev. D {\bf 68}, 093002 (2003),
hep-ph/0307323.

\bibitem{BHS} J.D. Bjorken, P. F. Harrison and W. G. Scott,
hep-ph/0511201.

\bibitem{MNY} R.~N.~Mohapatra, S.~Nasri and H.~B.~Yu,
  arXiv:hep-ph/0605020.


\bibitem{king} S.~F.~King, Nucl.\ Phys.\ B {\bf 576}, 85 (2000);
   S.~F.~King and N.~N.~Singh, Nucl.\ Phys.\ B {\bf 591}, 3 (2000).

\bibitem{ps}In addition to the papers just cited, earlier
references are S. Pakvasa and H. Sugawara,
Phys. Lett. B {\bf 73}, 61 (1978);
 {\bf 82}, 105 (1979); E. Derman and H.S.Tsao, Phys.
 Rev. D {\bf 20}, 1207 (1979) and Y. Yamanaka, H. Sugawara
and S. Pakvasa Phys. Rev. D {\bf 25}, 1895 (1982).

\bibitem{cfm}S.-L. Chen, M. Frigerio and E. Ma, hep-ph/0404084.

\bibitem{fty}M. Fukugita, M. Tanimoto and T. Yanagida,
Phys. Rev. D {\bf 57}, 4429 (1998), hep-ph/9709388.

\bibitem{mr}E. Ma and G. Rajasekaran, Phys. Rev. D {\bf 64},
113012 (2001), hep-ph/0106291.

\bibitem{gmo}M. Gell-Mann, Phys. Rev. {\bf 125}, 1067 (1962);
S. Okubo, Prog. Theor. Phys. {\bf 27}, 949 (1962); {\bf 28},
 24 (1962).

\bibitem{numassdifferences} M. Maltoni, T. Schwetz, M. A. Tortola and J. W. F. Valle,
Phys. Rev. D {\bf 68}, 113010 (2003). For the choice $m_2>m_1$ see
A. de Gouvea, A. Friedland and H. Murayama, Phys. Lett. {\bf B490},
125 (2000). For recent reviews see, for examples, S. Pakvasa and J.
W. F. Valle, special issue on neutrino, Proc. Indian Natl. Sci.
Acad. Part A {\bf 70}, 189 (2004); arXiv:hep-ph/0301061 and V.
Barger, D. Marfatia and K. Whisnant, Int. J. Mod. Phys. E {\bf 12},
569 (2003); arXiv:hep-ph/0308123 ; R. N. Mohapatra and A. Y.
Smirnov, to appear in Ann. Rev. Nucl. Science, {\bf 56} (2006).

\bibitem{cosmobound}D. N. Spergel et al, Astrophys. J. Suppl.
{\bf 148}: 175 (2003);
S. Hannestad, JCAP {\bf 0305}: 004 (2003).

\bibitem{nsm}S. Nasri, J. Schechter and S. Moussa, Phys. Rev. {\bf D70},
053005 (2004); S.S. Masood, S. Nasri and J. Schechter, Phys. Rev. D
{\bf 71}, 093005 (2005); Int. J. Mod. Phys. A {\bf 21}, 517 (2006).
Graphical representation of this information may be found in
S. M. Bilenky, C. Giunti, J.A. Grifols and E. Masso, Phys. Rep.
{\bf 379}, 69 (2003).

\bibitem{kka}H. V. Klapdor-Kleingrothaus {\it et al}, Eur. Phys. J. A
{\bf 12}, 147 (2001). See also the recent review, C. Aalseth
{\it et al}, arXiv:hep-ph/0412300.

\bibitem{mutau} T. Fukuyama and H. Nishiura, hep-ph/9702253;
R. N. Mohapatra and S. Nussinov, Phys. Rev. {\bf D 60}, 013002
(1999); E. Ma and M. Raidal, Phys. Rev. Lett. {\bf 87}, 011802
(2001); C. S. Lam, hep-ph/0104116; T. Kitabayashi and M. Yasue,
Phys.Rev. {\bf D67} 015006 (2003); W. Grimus and L. Lavoura,
hep-ph/0305046; 0309050; Y. Koide, Phys.Rev. {\bf D69}, 093001
(2004);Y. H. Ahn, Sin Kyu Kang, C. S. Kim, Jake Lee, hep-ph/0602160;
 A. Ghosal, hep-ph/0304090; W. Grimus and L. Lavoura, Phys.\ Lett.\ B {\bf 572}, 189 (2003);
W.~Grimus and L.~Lavoura, J.\ Phys.\ G {\bf 30}, 73 (2004).


\bibitem{moh}  W. Grimus, A. S.Joshipura, S. Kaneko, L.
Lavoura, H. Sawanaka, M. Tanimoto, hep-ph/0408123; R. N. Mohapatra,
JHEP, {\bf 0410}, 027 (2004); A. de Gouvea, Phys.Rev. {\bf D69},
093007 (2004); R. N. Mohapatra and W. Rodejohann, Phys. Rev. {\bf D
72}, 053001 (2005); T. Kitabayashi and M. Yasue, Phys. Lett,. {\bf B
621}, 133 (2005);  R.~N.~Mohapatra and S.~Nasri, Phys.\ Rev.\ D {\bf
71}, 033001 (2005);R.~N.~Mohapatra, S.~Nasri and H.~B.~Yu, Phys.\
Lett.\ B {\bf 615}, 231 (2005).

\bibitem{Gutmutau}  K. Matsuda and H. Nishiura,
 Phys.\ Rev.\ D {\bf 73}, 013008 (2006);  A. Joshipura, hep-ph/0512252;
R. N. Mohapatra, S. Nasri and H.~B.~Yu, Phys. Lett. {\bf B 636}, 114
(2006).



\end{thebibliography}
\end{document}